\documentclass{iau} 
\usepackage{graphicx}

\usepackage{wrapfig}

\usepackage{graphicx}
\usepackage{sidecap}
\usepackage{graphicx}
\usepackage{natbib}
\usepackage{multicol}
\usepackage{color}
\usepackage[usenames,dvipsnames]{xcolor}
\usepackage{url}
\usepackage{hyperref}
\usepackage{color}
\definecolor{darkblue}{rgb}{0.0, 0.0, 0.5}
\definecolor{darkred}{rgb}{0.8, 0.0, 0.0}

\hypersetup{colorlinks=true,citecolor=darkblue,linkcolor=darkblue}

\usepackage{setspace}

\def\aj{AJ}%
\def\apj{ApJ}%
\def\aap{A\&A}%
\def\mnras{MNRAS}%
\def\pasp{PASP}%
\renewenvironment{thebibliography}[1]{%
 \thebib@list
}{
 \endlist
}

\def\thebib@list{%
 \list{\null}{%
 \partopsep 0mm
 \leftmargin 1.2em
 \labelsep 0mm
 \itemindent -1.2em
 \itemsep 0.1\baselineskip
 \parsep 0mm
  \usecounter{enumi}%
 }%
}%

\title[What can the occult do for you?] 
{What can the occult do for you?}

\author[B.W. Holwerda]   
{B.W. Holwerda$^1$
 \and W.C. Keel$^2$}

\affiliation{
$^1$ Sterrenwacht, University of Leiden, Niels Bohrweg 2, NL-2333 CA Leiden, the Netherlands \\ email: {\tt holwerda@strw.leidenuniv.nl} twitter: {\tt @benneholwerda} \\[\affilskip]
$^2$ Dept. of Physics \& Astronomy, University of Alabama, Box 870324, Tuscaloosa, AL 35487, USA
 \\email: {\tt keel@ua.edu} twitter: {\tt @NGC3314}}

\pubyear{2016}
\volume{321}  
\jname{Formation and evolution of galaxy outskirts}
\editors{A. Gil de Paz, J. Knapen \& J. Lee, eds.}

\begin{document}

\maketitle

\begin{abstract}
 Interstellar dust is still a dominant uncertainty in Astronomy, limiting precision 
in e.g., cosmological distance estimates and models of how light is
re-processed within a galaxy. When a foreground galaxy serendipitously overlaps
a more distant one, the latter backlights the dusty structures in the nearer
foreground galaxy. 

Such an overlapping or occulting galaxy pair can be used to
measure the distribution of dust in the closest galaxy with great accuracy. The
STARSMOG program uses Hubble to map the distribution of dust in foreground 
galaxies in fine ($<$100 pc) detail. Integral Field Unit (IFU) observations will map the
effective extinction curve, disentangling the role of fine-scale geometry and 
grain composition on the path of light through a galaxy.

  The overlapping galaxy technique promises to deliver a clear understanding of
the dust in galaxies: geometry, a probability function of dimming as a function 
of galaxy mass and radius, its dependence on wavelength.

\end{abstract}

\firstsection 
\section{Introduction}

Dust remains an issue for the study of stellar populations in galaxies, not just within
\begin{wrapfigure} [17]{l}{0.49\textwidth}
\includegraphics [width=0.49\textwidth]{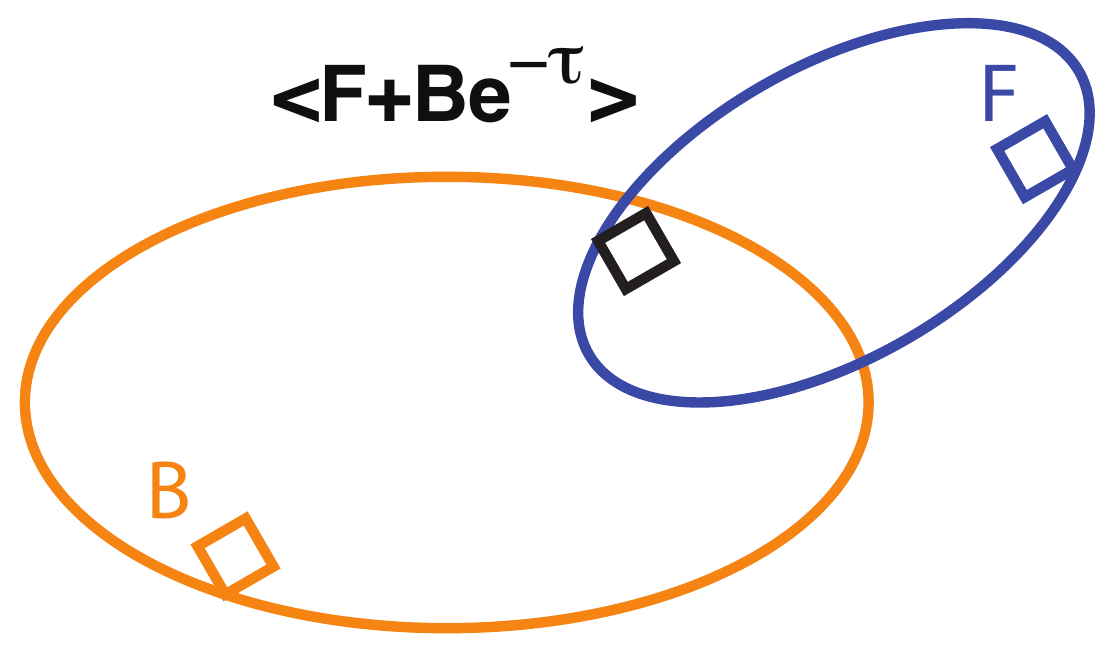}
\caption{Measuring dust in overlapping galaxies: dust extinction in the overlap region ({\bf black aperture}) can be estimated from the complementary apertures in the foreground spiral ({\color{Blue}F}) and the background galaxy ({\color{Orange}B}): $A_V = 1.086 \times \bar{\tau} = - 1.086 \times ln \left( {\bf (F+Be^{-\tau}) - {\color{ForestGreen} \bar{F}} \over {\color{red} \bar{B}}} \right)$ }
\end{wrapfigure}
the optical disk of a galaxy but outside it as well. To study dust in the outskirts of galaxies,
overlapping or occulting galaxy pairs are eminently suited. Unlike Herschel observations,
the occulting galaxy technique does not suffer from source confusion and it can measure
zero extinction, ruling out dust at a certain outer radius.

Using our much expanded sample of overlapping galaxies from SDSS, GAMA spectroscopy 
\citep{Holwerda07c,Holwerda15} and the GalaxyZoo \citep{Keel13}, totaling $\sim$ 2500 pairs, 
we can now start to address some of the dust topics in the outskirts of spiral galaxies:
To what radius is there still evidence for dust attenuation?
To what height above the plane is dust still present?
What is the distribution of extinction values --$P(A_V)$-- in the outskirts?
Is there a different attenuation law in the low-metallicity environment of the outer disk?

\begin{wrapfigure} [17]{l}{0.49\textwidth}
\includegraphics [width=0.49\textwidth]{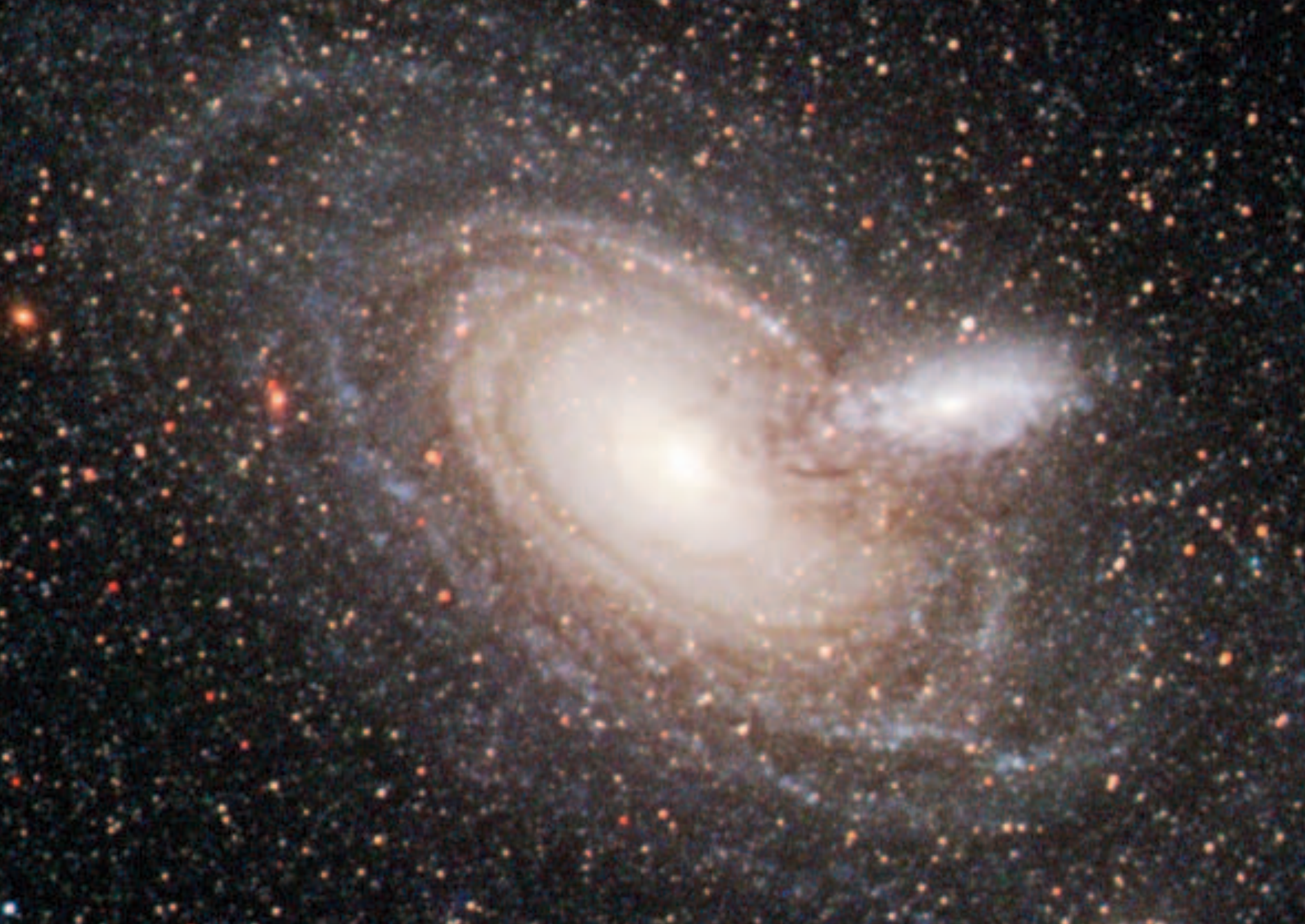}
\caption{\label{f:angstpair}The Hubble Heritage image of a $M^*\sim 10^9 M_\odot$ galaxy in overlap of a bright background, highlighting dust well into the outskirts \protect{\citep{Holwerda09}}.}
\end{wrapfigure}

The first question, the radial extent of dust in galaxies is clearly pointing to dust absorption still to be present in
galaxies well outside the typical optical disk ($R_{25}$ or the Petrosian radius). This appears to be true in both massive disks \citep[][]{kw00a,kw00b,kw01a,kw01b}, as well as smaller, low-mass disk galaxies \citep[][Lumbreras-Calle et al. {\em submitted}]{Holwerda09}.
Which leaves us with the issue of how this dusty ISM medium got there. Is it {\em in situ} formation, proportional to the star-formation in the outskirts? or has it been transported out in some type of radial flow (similar to stellar migration) or perhaps the product of recent accretion that is not as pristine as originally thought? It is likely a combination of all of the above. 
\begin{wrapfigure} [18]{r}{0.49\textwidth}
\includegraphics [width=0.49\textwidth]{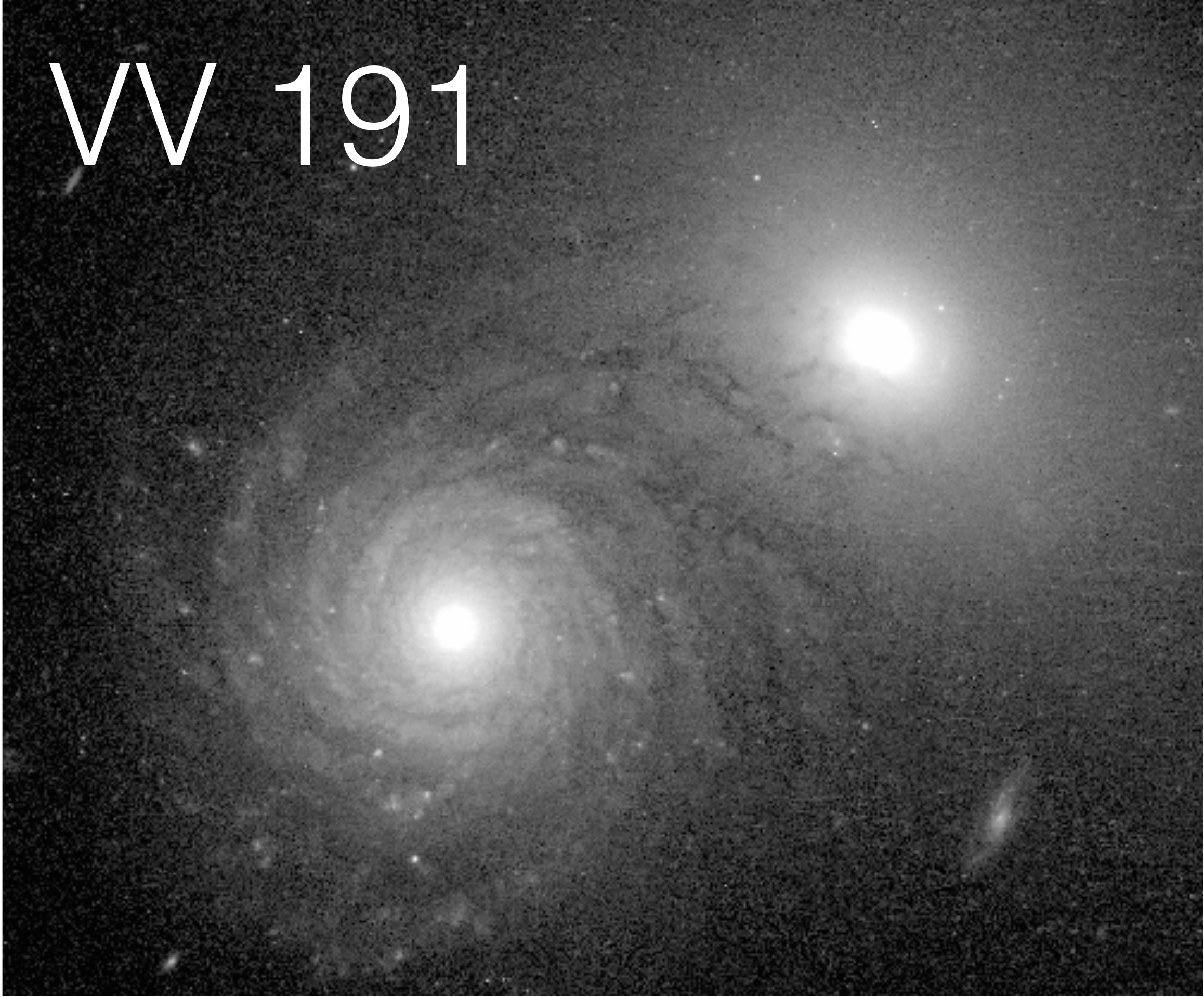}
\caption{VV191, recent example from STARSMOG. Dusty structure can be found extending throughout the disk of the foreground, face-on spiral. }
\end{wrapfigure}

For the overlapping galaxy technique remains to quantify the dust reservoir in the outskirts and the typical radial density profile. Evidence of both star-formation and stellar populations throughout the HI gas disk were shown in several presentation (e.g. Abrahams). With overlapping galaxies, it will be possible to constrain how the dust (and hence molecular) component is spread throughout (see Watson's talk on the difficulty of obtaining CO observations in the outer disk). The aim of the ongoing STARSMOG \citep[][]{Holwerda14} HST program is to map the distribution of attenuation values throughout the disks of nearby galaxies -- $P(A_V)$. The outskirts of galaxies have dust in them and the overlapping galaxies approach is poised to map how much and its distribution. 

\section{Reddening Law}

\begin{figure}[htbp]
	\includegraphics[width=0.5\textwidth]{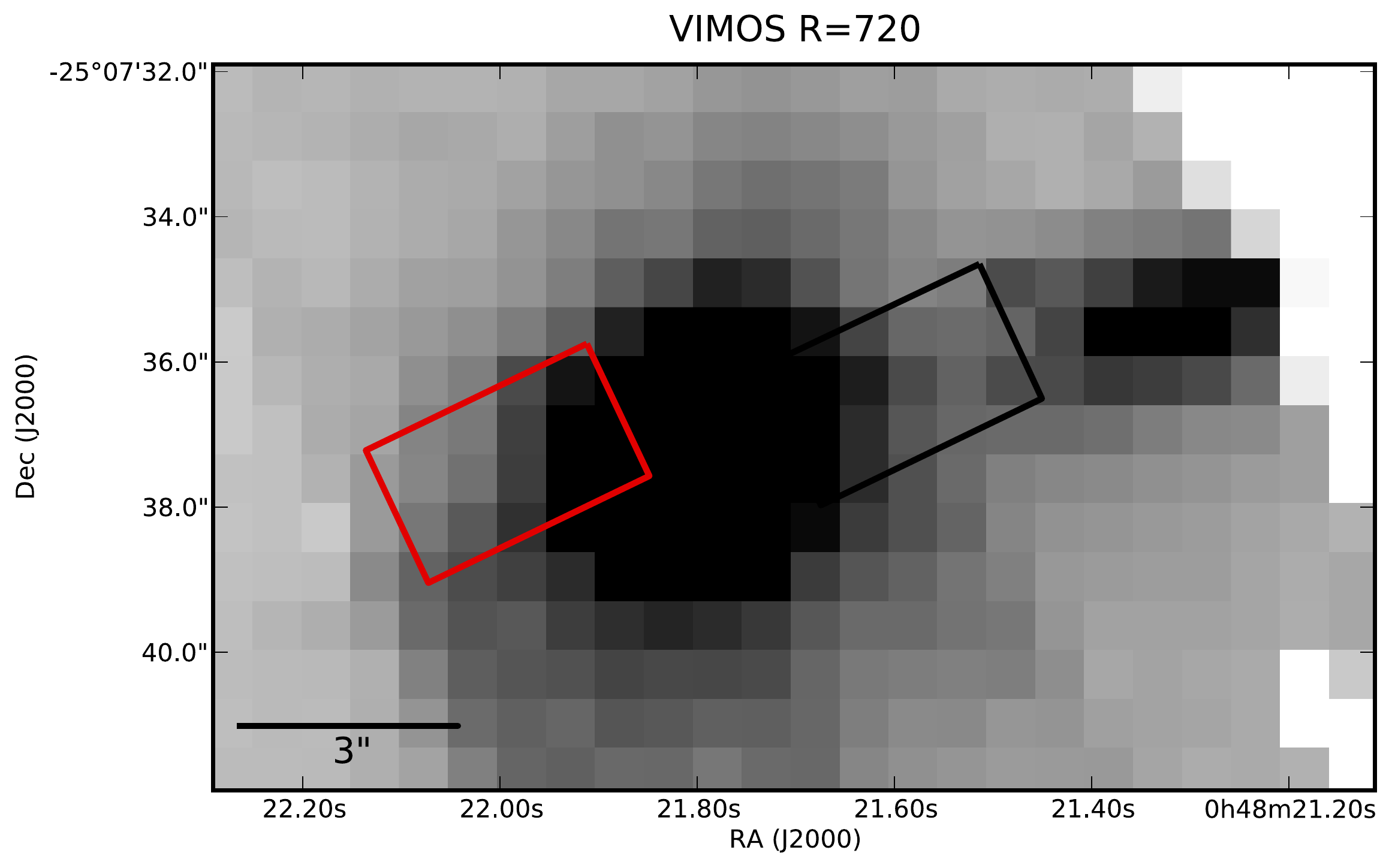}
	\includegraphics[width=0.5\textwidth]{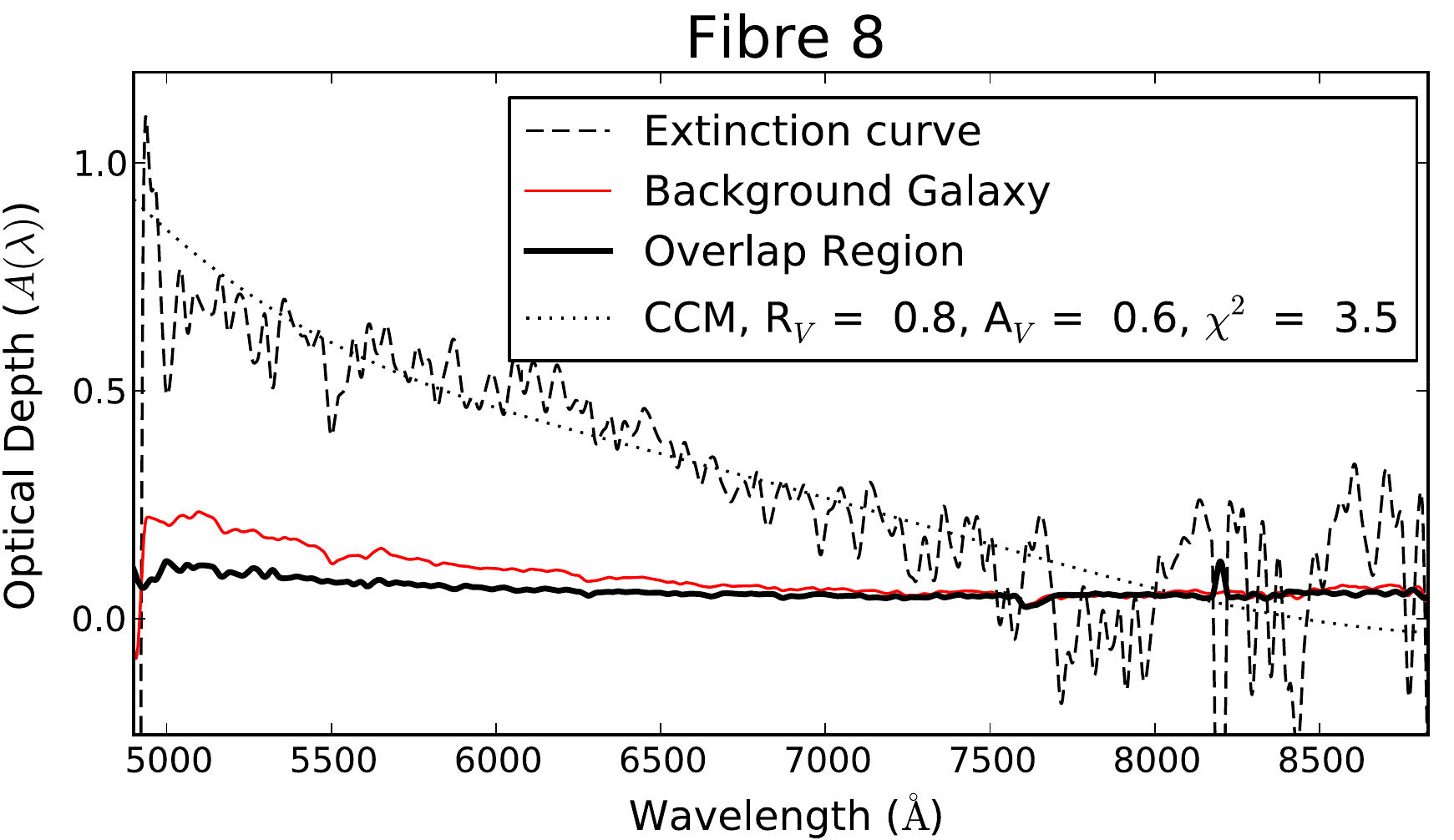}
	\caption{\label{f:vimos}VIMOS IFU observations of the pair in Fig. \ref{f:angstpair}. 
	To map the attenuation curve, each fibre spectra in the overlap region (black rectangle) is matched to a corresponding companion in the background galaxy (red rectangle). For each matched fibre pair, the attenuation curve (dashed line) can be fit with the \protect\cite{CCM} relation (dotted line).}
 \end{figure}

The dust attenuation can inferred from single filter images, the reddening-attenuation relation, i.e., the attenuation law. 
Based on the Balmer decrement (H$\alpha$/H$\beta$ ratio), several model attenuation laws have been put forward by \cite{Calzetti94,Calzetti99,Charlot00, Wild11}. 
Using either multiple broad-band filters \citep[e.g.][]{Holwerda09,Keel14} or IFU observations, one can map the attenuation curves in overlapping galaxy pairs. Pilot IFU observations \citep[][Fig. \ref{f:vimos}]{Holwerda13a,Holwerda13b} already reveal a variety of the reddening laws. 

Wavelength coverage and spatial resolution are critical to resolve the attenuation relation in nearby overlapping galaxy pairs. In case of multi wavelength observations, high-resolution ($\sim1$") ultraviolet imaging is critical to resolve the ``bump" at $\sim$2100 \AA\  in the reddening law. Alternative, high spatial resolution IFU can map reddening relations.

%

\vspace*{-0.7 cm}

\section{Outlook}

Future work on overlapping galaxies will focus on the analysis of large samples of pairs available from SDSS, GAMA, and other surveys. 
New imaging observations in the UV through near-infrared will map the attenuation-reddening relation (using two overlapping spirals) using, e.g., ASTROSAT,  HST and Spitzer.
IFU observations of select pairs (such as VV191 in Fig. ~1) will map the optical attenuation curve as a function of radius
and ISM fine structure. The ultimate goal of the STARSMOG project is to provide the extra-galactic community with probability functions
of both the amount of dust attenuation to be encountered -- $P(A_V)$ --  and the distribution of attenuation curves -- $P(R_V)$ --  to be used in the next generation of spectral energy distribution models of galaxies and distance measures with SNIa.
The typical dust content in the outer regions of spiral disks will be a valuable insight into the ``pristine" nature of the most recently added material and the star-formation of the outermost disk.

\vspace*{-0.2 cm}

 
%


\end{document}